\documentstyle[preprint,aps,amssymb,amsmath,epsfig]{revtex}

\newcommand{\dd}{{\rm d}}
\newcommand{\ee}{{\rm e}}

\begin{document}
\tighten

\title{D-Wave Superconductors near Surfaces and Interfaces:
A Scattering Matrix Approach within the Quasiclassical Technique}

\author{T. L\"uck and U. Eckern}
\address{Institut f\"ur Physik, Universit\"at Augsburg,
D-86135 Augsburg, Germany}

\author{A. Shelankov}
\address{Department of Theoretical Physics,
Ume{\aa} University, 901 87, Sweden\\
A.F. Ioffe Physico-Technical Institiute, 19021 St. Petersburg, Russia}

\date{\today}

\maketitle

\begin{abstract}
A recently developed method
[A. Shelankov and M. Ozana, Phys. Rev. B {\bf 61}, 7077 (2000)] is
applied to investigate $d$-wave superconductors in the vicinity of
(rough) surfaces. While this method allows the incorporation of
arbitrary interfaces into the quasiclassical technique, we discuss,
as examples, diffusive surfaces and boundaries with small tilted
mirrors (facets).  The properties of the surface enter via the
scattering matrix in the boundary condition for the quasiclassical
Green's function.  The diffusive surface is described by an ensemble
of random scattering matrices. 
We find that the fluctuations of the density of states around the average
are small; the zero bias conductance peak broadens with increasing 
disorder. The faceted surface is described in the model where the scattering
matrix couples $m$ in- and $m$ out-trajectories ($m \geq 2$).
No zero bias conductance peak is found for [100] surfaces; the relation to the
model of Fogelstr\"om {\it et al.} [Phys. Rev. Lett. {\bf 79}, 281 (1997)] is
discussed.

\end{abstract}


\section{Introduction}

The $d_{x^2-y^2}$ symmetry of the order parameter (OP) in high
temperature (high-$T_{\text{c}}$) superconductors is nowadays well established
by various phase-sensitive experiments,
the most striking being the observation of a half flux quantum
in a tricrystal geometry by Tsui {\it et al.}~\cite{Ts94}.
Also the zero-bias conductance peak (ZBCP) found in different tunnel
experiments~\cite{Ge88,Ma91,Le92,Ka95,Al97} on $[110]$-oriented boundaries
of YBa$_2$Cu$_3$O$_{7-\delta}$ gave a strong hint for $d$-wave symmetry
as was first pointed out by Hu~\cite{Hu94}. However, in more recent
experiments, several puzzling questions arose. For example a splitting of the
ZBCP was observed~\cite{Co97} and ZBCP's were reported even for
$[100]$-oriented junctions~\cite{Co97,We98}. Disorder effects were examined
as well in experiment where the disorder of the junctions was
increased by ion irradiation~\cite{Ap98}. A decreasing height of the ZBCP was
observed whereas the width stayed constant.

\begin{figure}
\epsfig{file=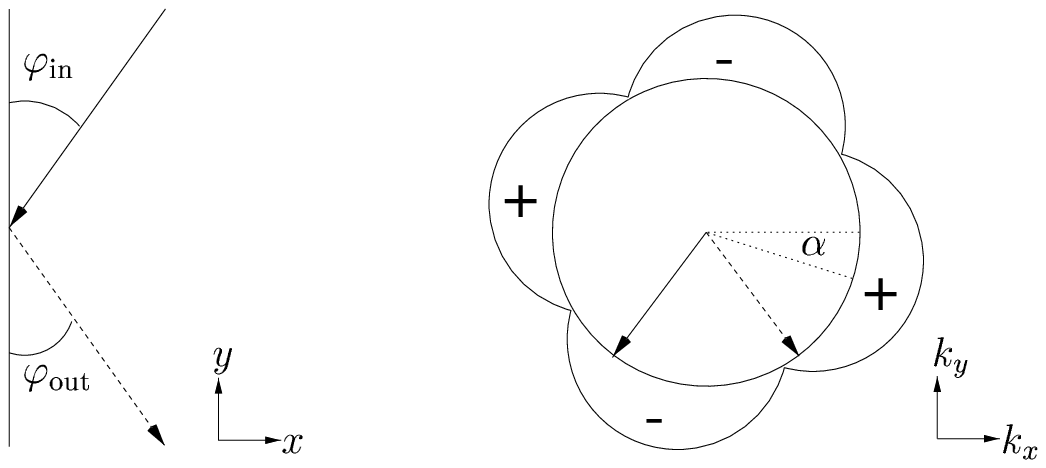}
\caption{Surface effect in real space (left) an k-space (right).}\label{spec}
\end{figure}

The most successful approach to treat such inhomogeneous
problems is the theory of quasiclassical Green's
functions~\cite{Ei68,Sc81,LaOv84}. The properties of surfaces
or interfaces are included by effective boundary condition.
For a specular
surface the Green's function has to be continuous on a classical 
trajectory (see Fig.~\ref{spec}).
In this most simple model the pair-breaking effect of surfaces as well as the
existence of a ZBCP can be explained: If quasiparticles are scattered to
branches with a different OP ($\alpha\neq 0$ ) the pairing is
suppressed and bound states can occur due to Andreev scattering. If a
quasiparticle is scattered with a sign change of the OP the ZBCP exists, too.
In this framework it was also shown~\cite{Ma95,Fo97} that the splitting of the
ZBCP is in agreement with the existence of a subdominant order parameter
($d_{x^2-y^2}+id_{xy}/s$) in the vicinity of the surface.
The generalization of the boundary conditions to a specular interface was
carried out by Zaitsev~\cite{Za84}. In this situation the Green's
function on four trajectories must match at the
interface, which leads to quite complicated non-linear boundary conditions.

As surface roughness is present in experiments, models were developed to
include disorder in the theory.
One possibility, first suggested by Ovchinnikov~\cite{Ov69}, is to
use the boundary conditions for the specular situation and cover the surface
with a thin dirty layer where equations for the dirty case must be applied.
In numerical studies of boundary problems similar techniques were
used~\cite{Bu97,Po00}.
Also a scattering matrix approach, which is related to the randomly rippled
wall model~\cite{Na96}, was applied to disordered surfaces~\cite{Ya96}; the
solution was given in a Born-like approximation. Except for models with
unitary scatterers~\cite{Po99}, all calculations show a broadening of the
ZBCP due to disorder.

Often the surface roughness is only present on scales much smaller than
the coherence length. In this case, the boundary conditions for the
quasiclassical Green's function can be formulated in terms of the scattering
matrix ($S$-matrix), as it has been recently suggested in~\cite{ShOz98}. 
A rough interface does not conserve the momentum parallel to the surface, and,
therefore, it couples waves ({\it i.e.}  quasiclassical trajectories) with
different propagation direction.
The microscopic structure of the interface enters the theory via the
$S$-matrix. In the absence of detailed knowledge about the microscopic
structure of the surface, the $S$-matrix has to be taken as a
phenomenological input.

\begin{figure}
\epsfig{file=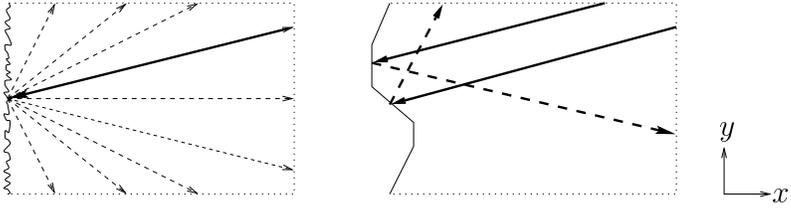}
\caption{Effect of roughness on small (left) and large (right).}
\label{coh-inc}
\end{figure}

Due to the very short in-plane coherence length in high-$T_{\text{c}}$
superconductors ($\xi_0\approx 20 \text{\AA}$) structures on larger
scales occur as well. 
Then the translational invariance parallel to the surface is lost on this
length scale and a full 2D treatment of the problem is necessary, in which
the trajectories are considered individually (see Fig.~\ref{coh-inc}).
For example, in cuprates facets with typical dimensions of $10-100 \text{nm}$
are present at interfaces~\cite{Ma96}. This leads to the existence of a ZBCP
even for [100] tunnel junctions as was pointed out by
Fogelstr\"om {\it et al.}~\cite{Fo97}.

In our study we use the scattering matrix approach presented in~\cite{ShOz98}
to describe surfaces with microscopic roughness.
We will discuss two kinds of surfaces: First, we study a
microscopically disordered surface, which is described by random matrices. In
contrast to earlier calculations we are able to consider individual
realizations of the disorder; we examine averaged quantities as well as
fluctuations.
Afterwards we focus on an $S$-matrix that describes a surface with
tilted mirrors, where few trajectories are connected at the surface.

In the following section we will briefly introduce the quasiclassical
theory for superconductors. Subsequently we will 
present the boundary conditions in the form most suitable for our purpose and
discuss the general properties. In section~\ref{ResSurf} we
derive the S-matrices for different situations and
present the related results. We conclude with a discussion of our results
and compare them to other approaches.

\section{Method}

\subsection{Theory of quasiclassical Green's functions}
In our studies of boundary effects of unconventional superconductors we use
the theory of quasiclassical Green's functions introduced by
Eilenberger~\cite{Ei68}.
This approximation of Gorkov's theory is valid in the quasiclassical
limit ($2\pi/k_{\rm F}\ll\xi$). Several review
articles have been written on this subject, e.g. by Schmid~\cite{Sc81} or
Larkin and Ovchinnikov~\cite{LaOv84}. The quasiclassical propagator in
Nambu space
\begin{equation}\label{gmat} 
\hat g(\omega,{\bf R},{\bf k}_{\rm F})=
\begin{pmatrix}
g & f\\f^{\dagger}& g^{\dagger}
\end{pmatrix}
\end{equation}
is determined by the Eilenberger equation
\begin{equation}\label{eile}
\left[\left(\omega+
\frac{e}{c}{\bf v}_{\bf F}\cdot{\bf A}({\bf R})\right)\hat\tau_3-
\hat\Delta({\bf R},{\bf k}_{\rm F}),
\hat g(\omega,{\bf R},{\bf k}_{\rm F})\right]+
i\hbar({\bf v_{\rm F}}\cdot{\boldsymbol\nabla})
\hat g(\omega,{\bf R},{\bf k}_{\rm F})=0
\end{equation}
where $\hat\tau_i$ represent the Pauli matrices and 
\begin{equation}\label{hatop}
\hat \Delta({\bf R},{\bf k}_{\rm F})=
\begin{pmatrix}
0 & \Delta\\ -\Delta^*& 0
\end{pmatrix}.
\end{equation}
Additionally a normalization condition is needed to obtain the physical
solution of the equation
\begin{equation}\label{norm}
\hat g^2(\omega,{\bf R},{\bf k}_{\rm F})=\hat 1.
\end{equation}
The OP must obey the self-consistency equation
\begin{equation}\label{selfcon}
\hat\Delta({\bf R},{\bf k}_{\rm F})=-
\int_{-\omega_{\rm c}}^{\omega_{\rm c}}\frac{\dd\omega}{4}
\left\langle V({\bf k}_{\rm F},{\bf k}'_{\rm F})
\hat g^{\rm K}(\omega,{\bf R},{\bf k}'_{\rm F})
\right\rangle_{{\bf k}'_{\rm F}}.
\end{equation}
Here $\langle\dots\rangle_{{\bf k}'_{\rm F}}$ indicates the average over the
Fermi surface. In thermal equilibrium the Keldysh propagator
$\hat g^{\rm K}$ is given by the the advanced and retarded propagator
$\hat g^{\rm A/R}$
 \begin{equation}\label{Keldysh}
\hat g^{\rm K}=(\hat g^{\rm R}-\hat g^{\rm A})\tanh(\omega/2k_{\rm B}T)
\qquad{\rm with}\qquad\hat g^{\rm R/A}: \omega\to\omega\pm i0_+.
\end{equation}
For simplicity we make some further assumptions concerning the microscopic
properties: For the interaction we choose
$V({\bf k}_{\rm F},{\bf k}'_{\rm F})=
V\cos[2(\varphi-\alpha)]\cos[2(\varphi'-\alpha)]$~\cite{Bu95}
which generates a $d$-wave OP
with orientation $\alpha$ (see Fig.~\ref{spec})
\begin{equation}
\Delta({\bf R},{\bf k}_{\rm F})=\Delta({\bf R})\cos[2(\varphi-\alpha)].
\end{equation}
In addition we assume
an isotropic two-dimensional model with a spherical Fermi surface.\\
After the determination of the OP all physical properties can be calculated
from the quasiclassical Green's function. For example, the angle-resolved
local density of states (DOS) reads
\begin{equation}\label{arDOS}
{\cal N}(\omega,{\bf R},{\bf k}_{\rm F})={\cal N}_0{\rm Re}
\left[
g^{\rm R}(\omega,{\bf R},{\bf k}_{\rm F})\right],
\end{equation}
where ${\cal N}_0$ is the normal state DOS.
In many cases the knowledge of the angle-averaged DOS is sufficient
\begin{equation}\label{avDOS}
{\cal N}(\omega,{\bf R})=\left\langle{\cal N}(\omega,{\bf R},{\bf k}_{\rm F})
\right\rangle_{{\bf k}_{\rm F}}.
\end{equation}
The DOS at the surface can directly be measured via the differential
conductance $G=\dd I/\dd V$ for normal-metal-insulator-superconductor
tunnel-junctions. For $T\to 0$ it is given by~\cite{Bu97}
\begin{equation}\label{Gdef}
G(V)=e^2A\langle v_{{\rm F},x}{\cal T}({\bf k}_{\rm F})
{\cal N}(eV,x=0,{\bf k}_{\rm F})
\rangle_{(k_{{\rm F},x}>0)},
\end{equation}
where the transmission probability is chosen as
\begin{equation}
{\cal T}(\varphi)=t^2\sin^2(\varphi)\ll 1,
\end{equation}
and $A$ is the area of the contact.
The current-density can be calculated from the Keldysh Green's
function via
\begin{equation}\label{current}
{\bf j}({\bf R})=-e{\cal N}_0\int\frac{\dd\omega}{4}
\left\langle
{\bf v}_{\rm F} {\rm Tr}
[\hat\tau_3\hat g^{\rm K}(\omega,{\bf R},{\bf k}_{\rm F})]
\right\rangle_{{\bf k}_{\rm F}}.
\end{equation}
It has been shown that the decomposition introduced by Maki
and Schopohl~\cite{MaSc95} is suitable for the numerical
integration of the Eilenberger equation as well as for analytical
considerations (see~\ref{sbcS})
\begin{equation}\label{decom}
\hat g=
\frac{1}{1-ab}
\begin{pmatrix}1+ab & -2a\\2b & -(1+ab)
\end{pmatrix}.
\end{equation}
Considering the physical meaning, the functions
$a(\omega,{\bf R},{\bf k}_{\rm F})$ and $b(\omega,{\bf R},{\bf k}_{\rm F})$
are closely related to the particle and hole amplitudes in the Andreev
equation as was discussed in detail in~\cite{ShOz98}.
With this construction the normalization condition is obeyed automatically.
By putting in this decomposition in Eq.~(\ref{eile}) it can be seen that 
the functions $a$ and $b$ are given by the equations
\begin{align}
\label{ria}
i\hbar({\bf v}_{\text{F}}\cdot{\boldsymbol\nabla}) a&=
\Delta^* a^2-2\omega a+\Delta,\\
\label{rib}
-i\hbar({\bf v}_{\text{F}}\cdot{\boldsymbol\nabla}) b&=
\Delta b^2-2\omega b+\Delta^*.
\end{align}
These equations can be solved on classical trajectories labeled by the
Fermi wave vector ${\bf k}_{\rm F}$. For each direction ${\bf v}_{\rm F}$
one has to integrate two ordinary differential equations in order to
construct the full propagator.\\
The Matsubara technique can be used as well to calculate the OP
\begin{equation}\label{Mat-selfcons}
\hat\Delta({\bf R},{\bf k}_{\rm F})=-
k_{\rm B}T\pi i\sum_{|\omega_n|<\omega_{\rm c}}
\left\langle V({\bf k}_{\rm F},{\bf k}'_{\rm F})
\hat f^{\rm M}(\omega_n,{\bf R},{\bf k}'_{\rm F})
\right\rangle_{{\bf k}'_{\rm F}}
\end{equation}
and the current-density
\begin{equation}\label{mat-current}
{\bf j}({\bf R})=-e{\cal N}_0 k_{\rm B}T\pi i\sum_{\omega_n}
\left\langle
{\bf v}_{\rm F}{\rm Tr}
[\hat\tau_3\hat g^{\rm M}(\omega_n,{\bf R},{\bf k}_{\rm F})]
\right\rangle_{{\bf k}_{\rm F}}.
\end{equation}
The energy integrals turn to sums over discrete Matsubara frequencies
$\omega_n=k_{\rm B}T\pi(2n+1)$ and the Matsubara propagator $\hat g^{\rm M}$
is determined by the relation
\begin{equation}
\hat g^{\rm R/A}(\omega,{\bf R},{\bf k}_{\rm F})=
\hat g^{\rm M}(\omega_n,{\bf R},{\bf k}_{\rm F})
\big|_{i\omega_n\to\omega\pm0_+}.
\end{equation}
One crucial point for investigating the effects of boundaries is still
missing. As
the quasiclassical condition does not apply in the vicinity of surfaces and
interfaces we have to treat the scattering of quasiparticles  by effective
boundary conditions. The properties of the boundary enter the calculations
only at this point.

\subsection{Boundary Conditions}\label{sbcS}

In our work we use the general theory recently derived in~\cite{ShOz98}. The
starting point is the Andreev-like equation
for the particle- and hole-like amplitudes which factorize the
Eilenberger Green's function in Eq. \ref{gmat} (see~\cite{ShOz98} for
details).  In this approach, it is possible to consider  roughness that
occurs on length scales much smaller than
the coherence length.  All information on the microscopic shape of the
boundary is provided by the scattering amplitudes from the
in-trajectories ($k^{\rm in}_{{\rm F},x}<0$) to the out-trajectories
($k^{\rm out}_{{\rm F},x}>0$); they are gathered in the scattering
matrix ${\bf S}$.

For simplicity we consider only a finite number $n$ of discrete in- and
out-trajectories
${\bf k}_{\rm F}^{\rm in/out}\to{\bf k}_{{\rm F},i}^{\rm in/out}$,
$i=1,2\ldots,n$ with equidistant angles.
Following~\cite{ShOz98} the boundary conditions are determined
using the functions
\begin{align}\label{bc-starta}
A_l(\beta)&=\det[{\bf 1}-{\bf S}\hat a{\bf S}^{\dagger}\hat b_l^\beta],\\
\label{bc-startb}
B_l(\alpha)&=\det[{\bf 1}-{\bf S}\hat a_l^\alpha{\bf S}^{\dagger}\hat b]
\end{align}
with the diagonal $n\times n$-matrices
\begin{equation}\label{bc-def}
\begin{split}
&\hat{a}= {\rm diag}\{a_{1},\dots, a_{n}\},\qquad  
\hat{a}_{l}^\alpha=
{\rm diag}\{a_{1},\dots ,a_{l-1}, \alpha , a_{l+1}, \dots\},\\
&\hat{b}= {\rm diag}\{b_{1},\dots, b_{n}\},\qquad  
\hat{b}_{l}^\beta= 
{\rm diag}\{b_{1},\dots, b_{l-1}, \beta , b_{l+1}, \dots\},\\
&\text{and}\qquad
a_{i} =a(\omega,x=0,{\bf k}_{{\rm F},i}^{\rm in}), \qquad 
b_{i} =b(\omega,x=0,{\bf k}_{{\rm F},i}^{\rm out}).
\end{split}
\end{equation}
The solutions of $A_l(\beta)=0$ and $B_l(\alpha)=0$ provide the boundary
conditions
\begin{align}\label{bc-solvea}
&A_l(\beta_0)=0\quad\Rightarrow\quad
a(\omega,x=0,{\bf k}_{{\rm F},l}^{\rm out})=\frac{1}{\beta_0},\\
\label{bc-solveb}
&B_l(\alpha_0)=0\quad\Rightarrow\quad
b(\omega,x=0,{\bf k}_{{\rm F},l}^{\rm in})=\frac{1}{\alpha_0}.
\end{align}
As the determinant is a linear function of each of the matrix elements the
functions $A_l(\beta)$ and $B_l(\alpha)$ are linear in $\beta$ and $\alpha$.
We can solve the boundary condition by calculating $A_l(\beta)$ and $B_l(\alpha)$
for two arbitrary values of $\beta$ and $\alpha$; for
$\beta = 0,1$ and $\alpha=0,1$ we obtain 
\begin{align}\label{bc-pm1}
a(\omega,x=0,{\bf k}_{{\rm F},l}^{\rm out})&=
1-\frac{A_l(1)}{A_l(0)},\\
\label{bc-pm2}
b(\omega,x=0,{\bf k}_{{\rm F},l}^{\rm in})&=
1-\frac{B_l(1)}{B_l(0)}.
\end{align}
With the boundary condition the Green's function can be calculated: At first
the integration of Eq.~(\ref{ria}) on the in- and
of Eq.~(\ref{rib}) on the out-trajectories is performed starting from the
known bulk values~\cite{ShOz98}
\begin{align}\label{bulk-starta}
a(\omega, x\to\infty, {\bf k}_{\rm F}^{\rm in})&=
\frac{\Delta_\infty({\bf k}_{\rm F}^{\rm in})}
{\omega+\sqrt{\omega^2-|\Delta_\infty({\bf k}_{\rm F}^{\rm in})|^2}},\\
\label{bulk-startb}
b(\omega, x\to\infty, {\bf k}_{\rm F}^{\rm out})&=
\frac{\Delta^*_\infty({\bf k}_{\rm F}^{\rm out})}
{\omega+\sqrt{\omega^2-|\Delta_\infty({\bf k}_{\rm F}^{\rm out})|^2}}
\end{align}
towards the boundary ($\Delta_\infty$: bulk OP).
Then the boundary conditions must be
applied to get the $a$'s on the out- and the $b$'s on the in-trajectories
at the boundary and the succeeding integration on these trajectories provides
the missing $a$'s and $b$'s.

The properties of the boundaries enter only via the $S$-matrix. The value
$|S_{ij}|^2$ is the probability of scattering from
${\bf k}_{{\rm F},j}^{\rm in}$ to ${\bf k}_{{\rm F},i}^{\rm out}$. We choose
the numbering of the trajectories so that ${\bf S}={\bf 1}$
reproduces the specular case. Due to current conservation ${\bf S}$ must be
unitary
\begin{equation}\label{unitarity}
{\bf SS}^{\dagger}={\bf 1}.
\end{equation}
With a suitable choice of ${\bf S}$ arbitrary physical situations can be
treated by this technique. Some examples are presented in
chapter~\ref{ResSurf}.

\begin{figure}
\epsfig{file=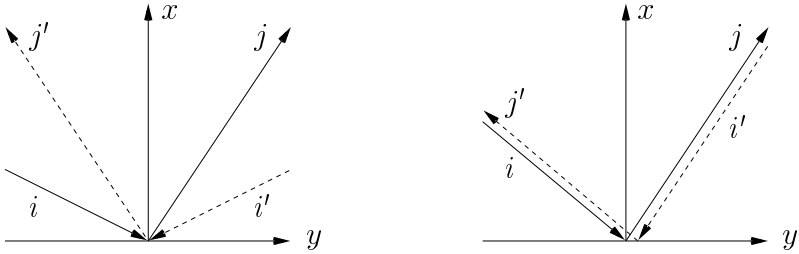}
\caption{Two trajectories (solid and dashed line) are plotted which are
connected via symmetry transformation for mirror (left) and time-reversal
symmetry (right).}\label{Symms}
\end{figure}

We are also able to connect basic symmetries of the physical situation with
transformation properties of the scattering matrix. The symmetry operations
for the mirror and the time-reversal symmetry are illustrated in
Fig.~\ref{Symms}. The mirror symmetry of the surface ($y\to-y$) is given by
the transformation
\begin{equation}\label{sym-mirror}
{\bf S}'={\bf TST},\quad\text{with}\quad T_{ij}=\delta_{(n+1-i),j}.
\end{equation}
The time-reversal symmetry operation is represented by
\begin{equation}\label{sym-tr}
{\bf S}'={\bf T}{\bf S}^T{\bf T}.
\end{equation}

\section{Results for different surfaces}\label{ResSurf}

Since we are discussing different types of roughness which occur in
experiments we have to find adequate scattering matrices for each situation.
As the unitarity condition~(\ref{unitarity}) must be obeyed we represent
${\bf S}$ by the relation
\begin{equation}\label{expiH}
{\bf S}=\exp\{i{\bf H}\}\qquad\text{with}\quad {\bf H}={\bf H}^\dagger.
\end{equation}
In the subsequent sections~\ref{RanSurf} and ~\ref{FacSurf} we present
S-matrices for random surfaces as well as for surfaces with small
tilted mirrors and
physical properties such as the OP and the DOS in the vicinity of a surface.
The retarded Green's-function should be evaluated at $\omega\to\omega+i\delta$;
for numerical purposes we keep $\delta$ finite.For the calculation
of the DOS we choose $\delta=0.02k_{\rm B}T_{\rm c}$ and $n=200$; we checked
that the results do not change for larger $n$.

\subsection{Random surface}\label{RanSurf}

In this section we search for a scattering matrix that can describe random
surfaces. To take into account the statistical properties of
the surfaces we choose a random matrix in this approach; therefore the
Hermitian matrix ${\bf H}$ is assumed to be a member of the Gaussian unitary
ensemble (GUE) as was also suggested by Yamada {\it et al.}~\cite{Ya96}
\begin{equation}\label{H-ran}
\begin{split}
&\langle H_{ij}\rangle=0,\\
&\langle H^*_{ij}H_{i'j'}\rangle=\frac{\tau}{n}\delta_{ii'}\delta_{jj'}.
\end{split}
\end{equation}
The brackets $\langle\dots\rangle$ denote the ensemble average of the disorder.
The roughness of the surface can be varied by the parameter $\tau$. The factor
$1/n$ in the correlator of ${\bf H}$ ensures that the whole procedure
does not depend on the number of channels for $n\to\infty$.

\begin{figure}
\epsfig{file=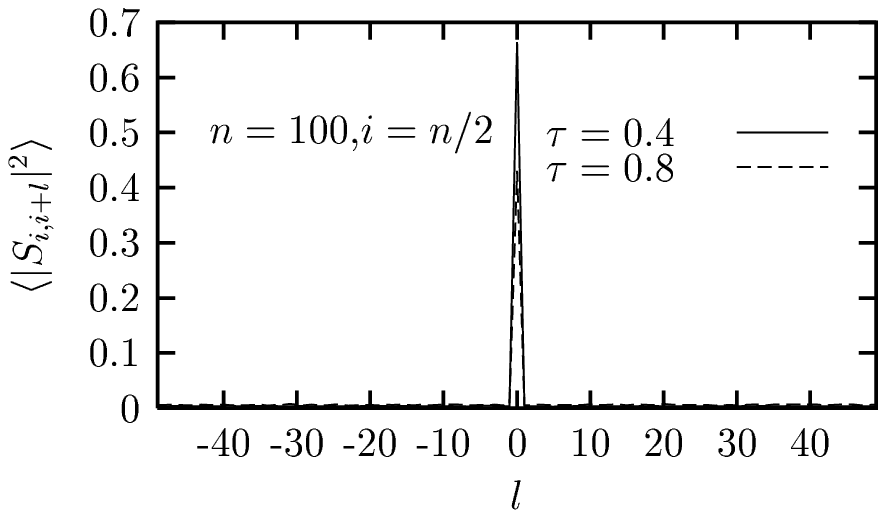}
\caption{Mean scattering probability from the in-trajectories $(i+l)$ to
a fixed out-trajectory $i=n/2$ for different values of $\tau$. The
specular contribution is reduced by increasing $\tau$.}\label{S.ran}
\end{figure}

\begin{figure}
\epsfig{file=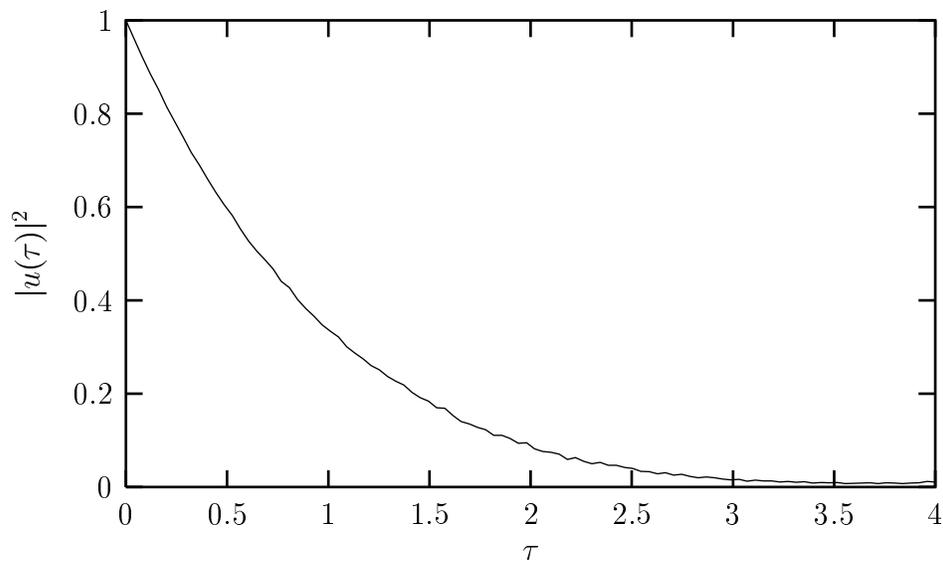}
\caption{Specular scattering weight as a function of $\tau$. The diffusive
limit is reached for $\tau\gtrsim2$.}\label{dif}
\end{figure}

The averaged scattering probabilities $\langle|S_{ij}|^2\rangle$ have a simple
behavior. For $\tau=0$ only $|S_{ii}|^2=1$ are finite and all other elements
are zero.
If $\tau$ is increased the diagonal elements (responsible for specular
reflection) are reduced to
$\langle|S_{ii}|^2\rangle=|u(\tau)|^2<1$ and
the off-diagonal elements become finite
$\langle|S_{i\neq j}|^2\rangle=|v(\tau)|^2\lesssim 1/n$
(see Fig.~\ref{S.ran}) with
\begin{equation}\label{relSii}
|u(\tau)|^2+(n-1)|v(\tau)|^2=1.
\end{equation}
Therefore, the averaged properties of ${\bf S}$ are fully determined by the
probability for specular reflection $|u|^2$ and we can use it as a measure
for the disorder of the
surface; its relation to the parameter $\tau$ is shown in Fig.~\ref{dif}.
For $\tau$ small enough, the reflection is partially specular. When 
$\tau\gtrsim 2$, the scattering becomes isotropic
since $|u|^2\sim |v|^2\sim 1/n$. We call this situation the diffusive limit.

\begin{figure}
\epsfig{file=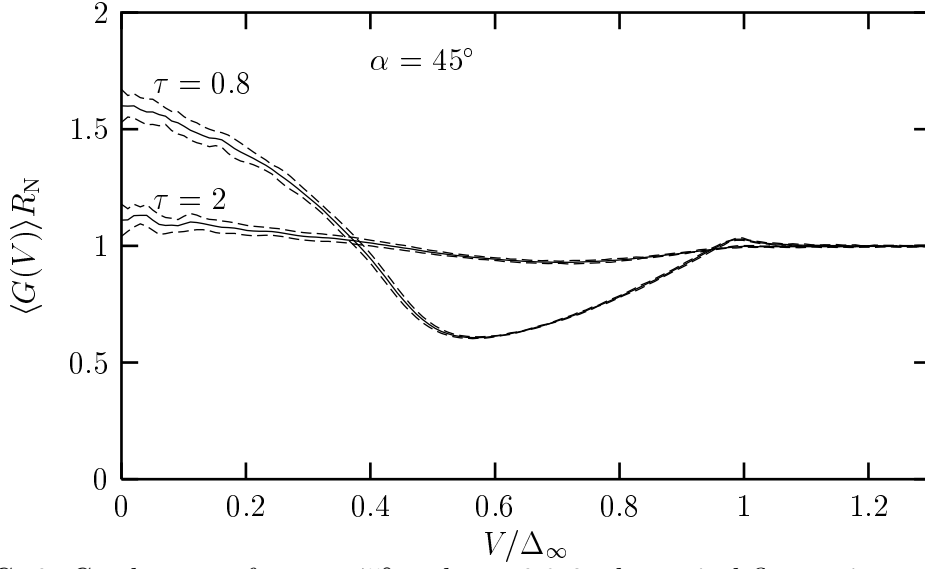}
\caption{Conductance for $\alpha=45^\circ$ and $\tau=0.8,2$; the typical
fluctuations are confined by the dashed lines.}\label{mdev}
\end{figure}

\begin{figure}
\epsfig{file=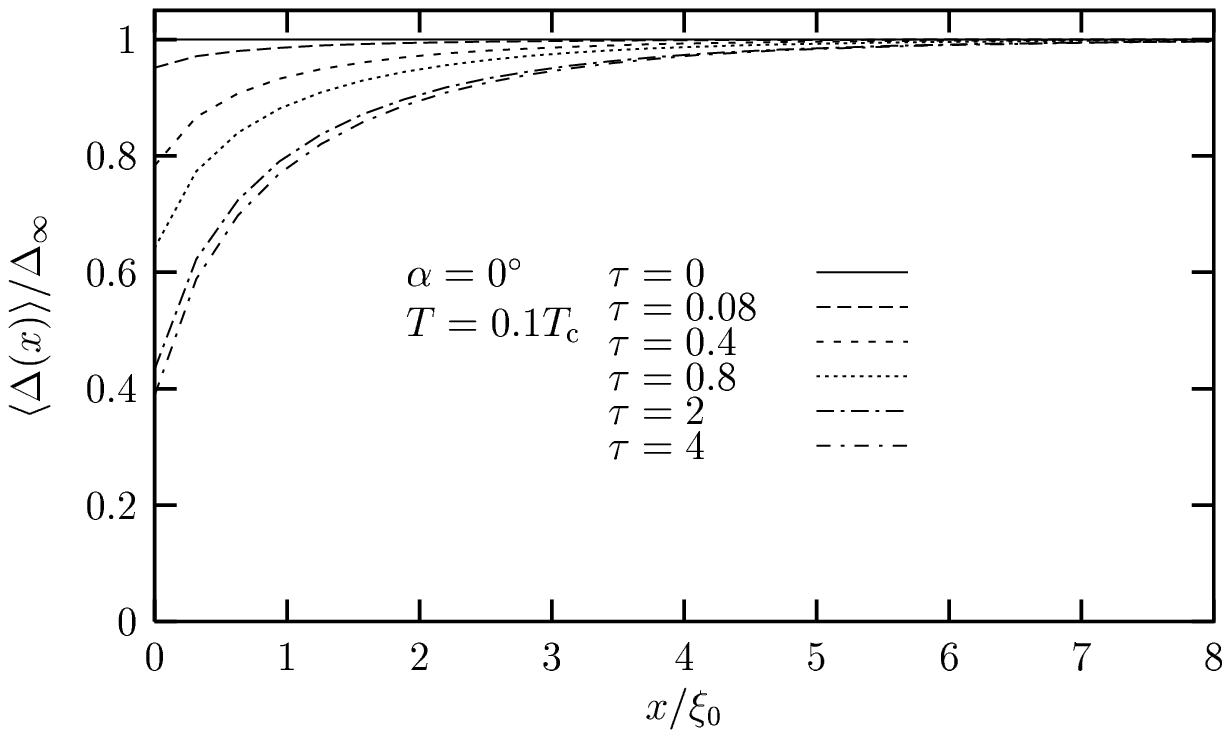}
\caption{Averaged order parameter for $\alpha=0^\circ$ and
$\tau=0,0.08,0.4,0.8,2,4$.}
\label{op0.ran}
\end{figure}

\begin{figure}
\epsfig{file=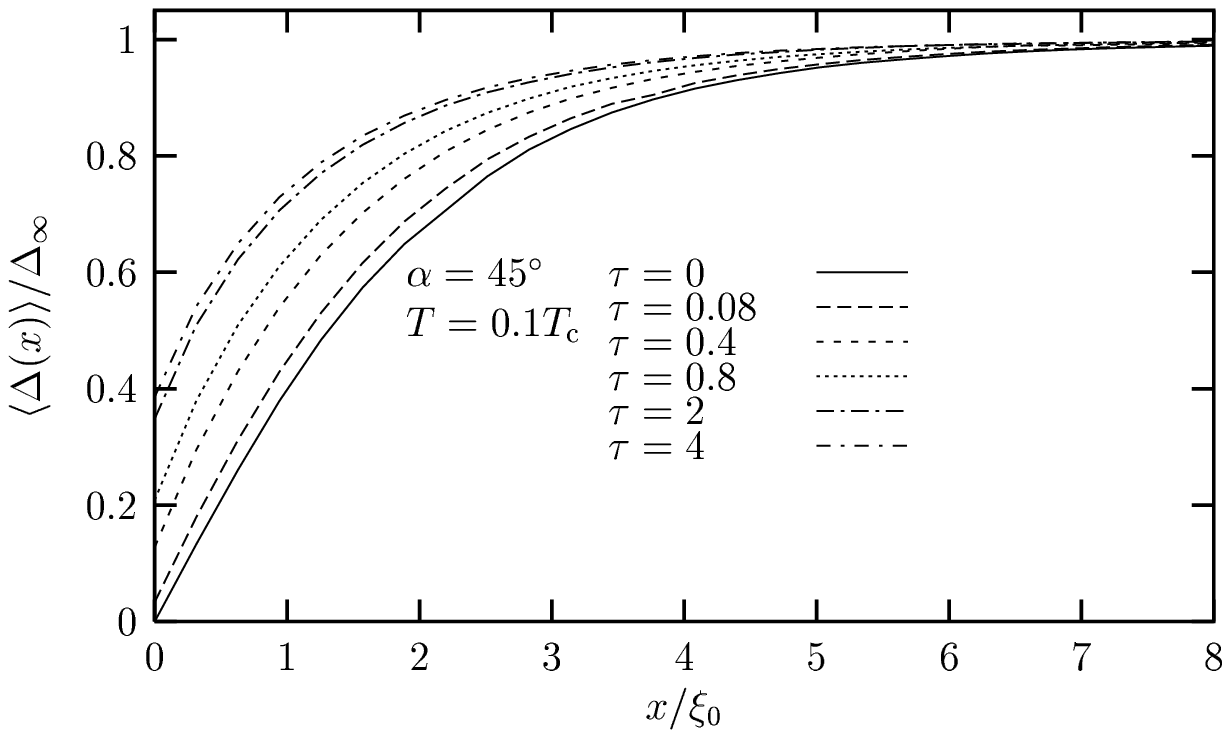}
\caption{Averaged order parameter for $\alpha=45^\circ$ and
$\tau=0,0.08,0.4,0.8,2,4$.}
\label{op45.ran}
\end{figure}

\begin{figure}
\epsfig{file=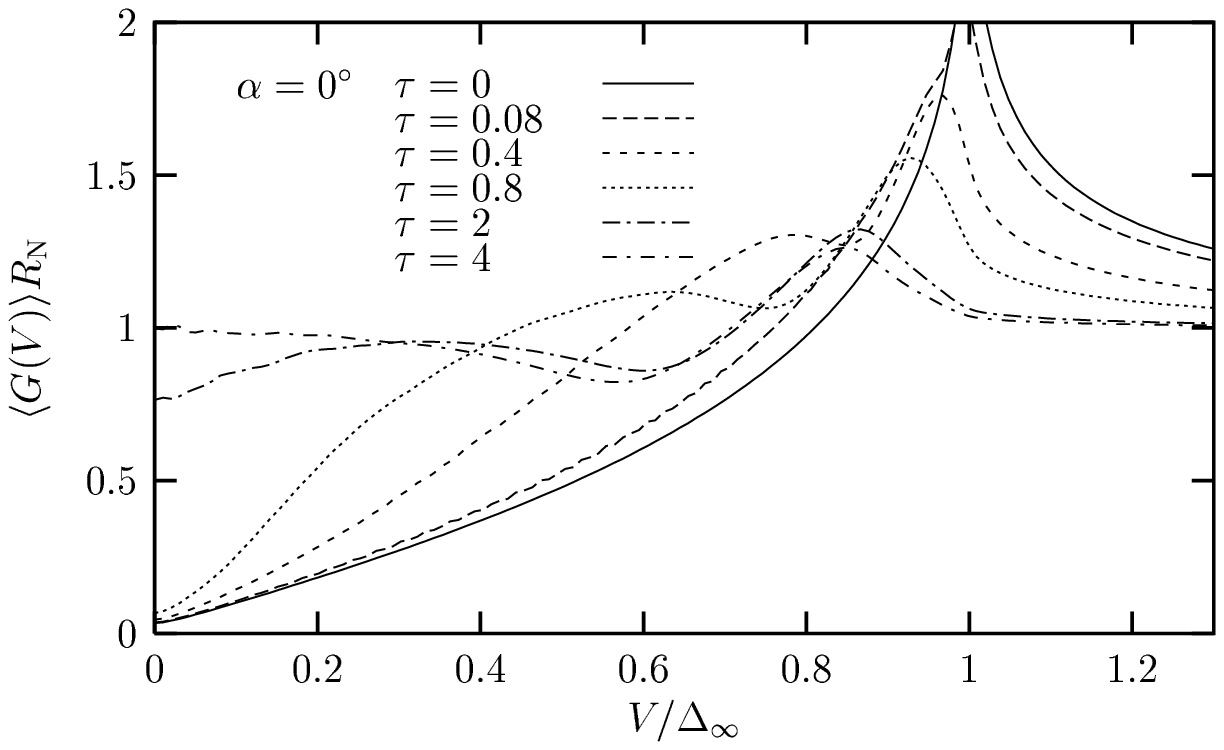}
\caption{Averaged differential conductance for $\alpha=0^\circ$ and
$\tau=0,0.08,0.4,0.8,2,4$.}
\label{dos0.ran}
\end{figure}

\begin{figure}
\epsfig{file=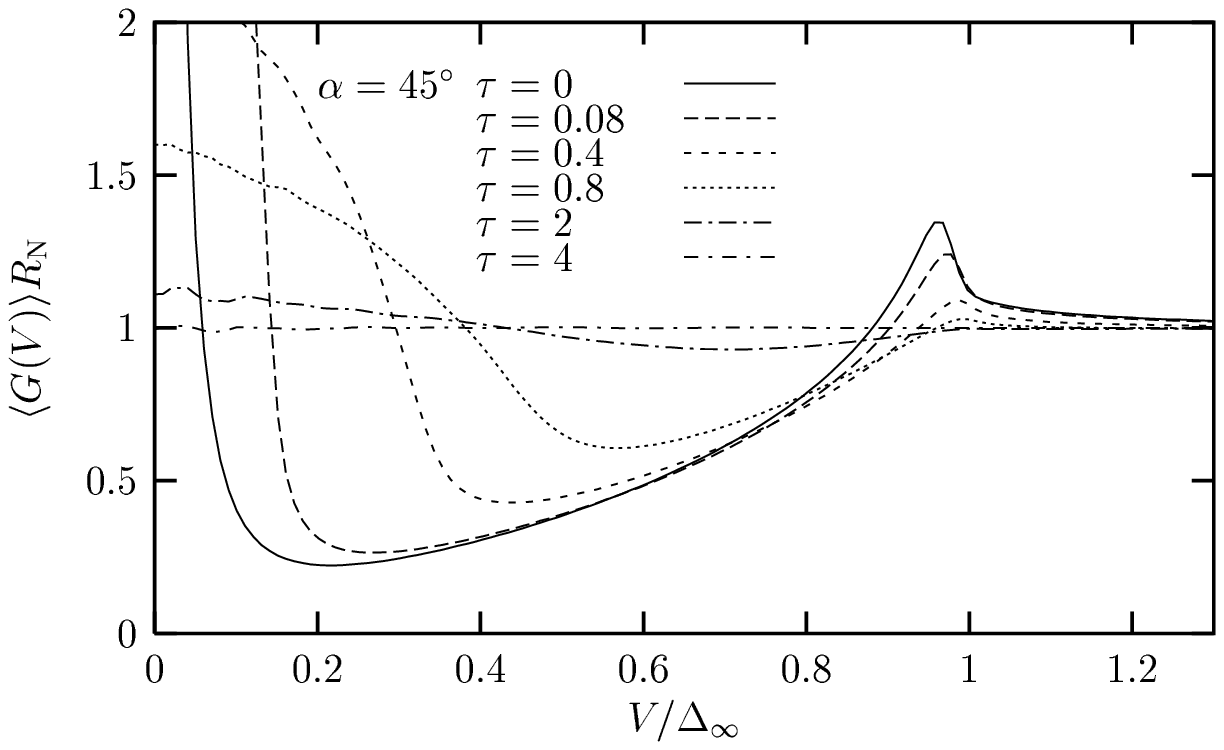}
\caption{Averaged differential conductance for $\alpha=45^\circ$ and
$\tau=0,0.08,0.4,0.8,2,4$.}
\label{dos45.ran}
\end{figure}

\begin{figure}
\epsfig{file=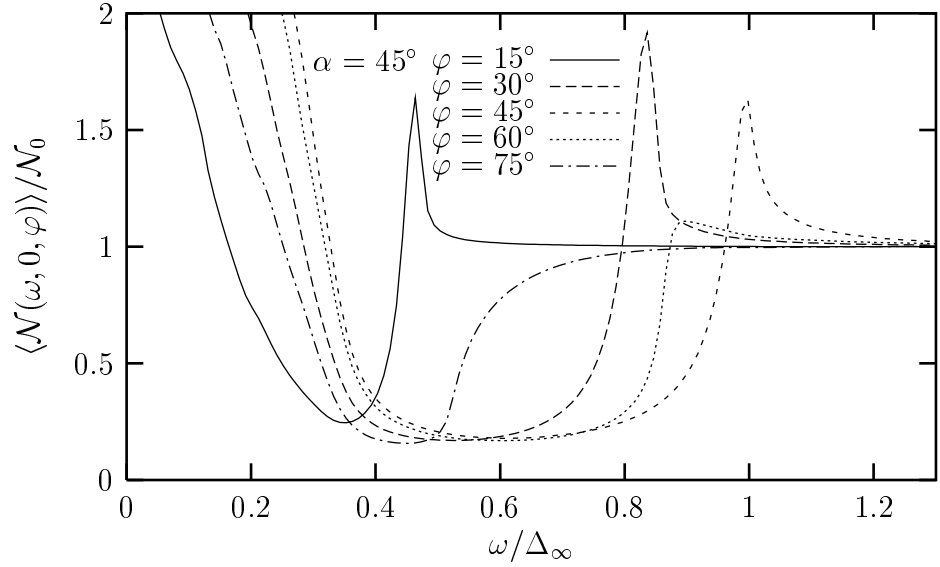}
\caption{Averaged angle-resolved DOS for $\alpha=45^\circ$ and $\tau=0.4$.}
\label{ados45.ran}
\end{figure}

We apply a random matrix ${\bf S}$ to calculate the OP and the DOS in the
vicinity of a disordered surface. We study several (up to 50) realizations of
the $S$-matrix individually. We find that the fluctuations are small
as can be seen in Fig.~\ref{mdev}. Therefore, it is meaningful to consider the
averaged quantities $\langle\Delta\rangle$ and $\langle G\rangle$.
The results for $\alpha=0^\circ,45^\circ$ and different
roughness values are shown in figs.~\ref{op0.ran} and~\ref{op45.ran} for the
OP and for the differential conductance in figs.~\ref{dos0.ran}
and~\ref{dos45.ran}, where the normal state resistance is used
\begin{equation}
R_N^{-1}=e^2A{\cal{N}}_0 t^2v_{\text{F}}\frac{4}{3\pi}.
\end{equation}
Additionally the angle-dependent DOS is presented
in Fig.~\ref{ados45.ran} for the medium roughness $\tau=0.4$.

We point out some interesting features of the model:
Disorder leads to a suppression and broadening of
the ZBCP for  $\alpha=45^\circ$. By comparison with experimental
data~\cite{Ap98} realistic results can be achieved by
$0.8\lesssim\tau\lesssim 2$. In our model with disorder,
for $\alpha=0^\circ$ no ZBCP occurs. In the angle-resolved DOS no splitting of
any bound states due to disorder is seen.
In the diffusive limit the OP reaches an almost
universal curve independent of the surface orientation $\alpha$; the
conductance
becomes flat and is of the order of the normal state value for all energies
and surface orientations.

\subsection{Surface with small tilted mirrors}\label{FacSurf}

\begin{figure}
\epsfig{file=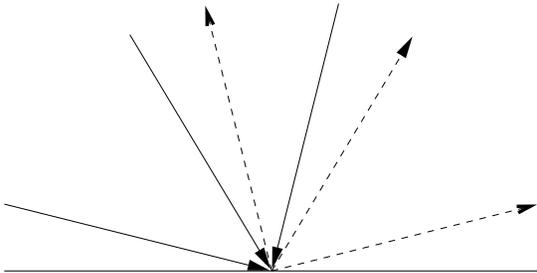}
\caption{For $m=3$ each of the three in-trajectories (solid lines) contributes
to the same three out-trajectories (dashed lines).}
\label{3facet-in-out}
\end{figure}

With this method also other roughness types can be studied.  In this section we
consider surfaces which only connect few trajectories.
For example one can think of small (compared to the coherence length) mirrors
with distinct orientations.  Each of the mirrors contributes to
reflection so that the surface acts as a beam-splitter. 
For the mirror orientations  $\theta=180^\circ l/m$ with
integer $l=-m+1,\dots,m-1$,  this can be achieved by
the choice
\begin{equation}\label{H-fac}
{\bf H}=\tau \tilde{\bf H}\qquad\text{with}\qquad
{\tilde H}_{ij}=\sum\limits_{l=1}^{m-1}\delta_{|i-j|,ln/m}.
\end{equation}
In the further calculation we choose $n$ to be a multiple of $m$; this leads
to a simpler form of ${\bf S}$ but has no physical implication for $n\gg m$.
In this case $\tilde{\bf H}$ considered as a block $m\times m$ matrix
has zeros on the diagonal,  and all other elements are equal to
${\bf 1}_{n/m}$ ($(n/m)\times (n/m)$ unity matrix). 
Since
\begin{equation}\label{kjb}
\tilde{\bf H}^2=(m-1){\bf 1}+(m-2)\tilde{\bf H}
\end{equation}
the exponential representation for the $S$-matrix in
Eq.~(\ref{expiH}) becomes
\begin{equation}\label{S-fac}
{\bf S}=u(\tau){\bf 1}+v(\tau)\tilde{\bf H},
\end{equation}
where
\begin{align}\label{unit-cond1}
&|u|^2+(m-1)|v|^2=1,\\
\label{unit-cond2}
&uv^*+u^*v+(m-2)|v|^2=0
\end{align}
as required by the unitarity of the $S$-matrix. For fixed $\tau$ the value of
$|u|^2$ is the probability for
specular reflection, whereas $|v|^2$ is the probability for the scattering on
one of the tilted mirrors. The scattering matrix can be calculated explicitly
and the amplitudes are given by
\begin{align}\label{scat-ampu}
&u(\tau)=\frac{\ee^{-i\tau}}{m}(m-1+\ee^{im\tau}),\\
\label{scat-ampv}
&v(\tau)=\frac{\ee^{-i\tau}}{m}(\ee^{im\tau}-1).
\end{align}
Therefore the probability for specular reflection (mirror with $\theta=0$) is
given by
\begin{equation}\label{scat-prob}
|u(\tau)|^2=\frac{1}{m^2}[(m-1)^2+1+2(m-1)\cos (m\tau)].
\end{equation}
In  this model, each incoming  trajectory is split  into  $m$ outgoing
trajectories. Altogether, there  are  $m$  in-trajectories which are
scattered into the same out-states. In the terminology of~\cite{ShOz98}, this
corresponds to a {\it knot}  with $m$ in- and
$m$ out-trajectories (see Fig.~\ref{3facet-in-out}).

\begin{figure}
\epsfig{file=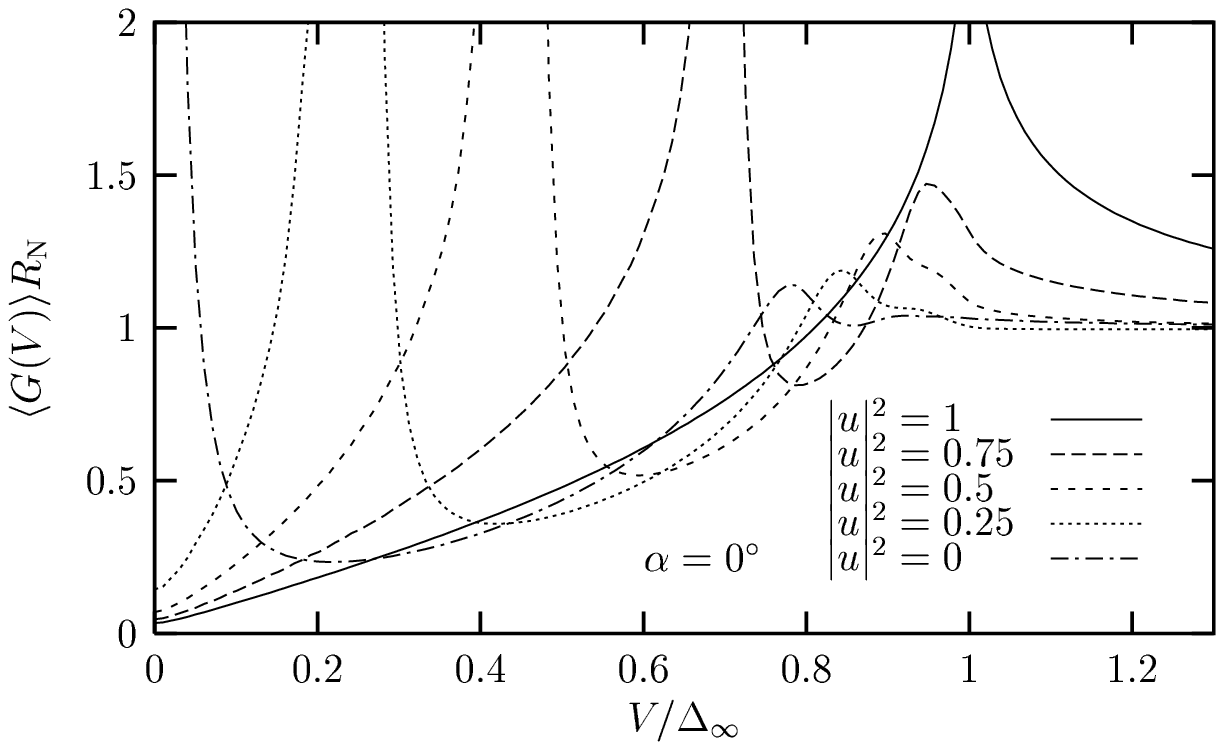}
\epsfig{file=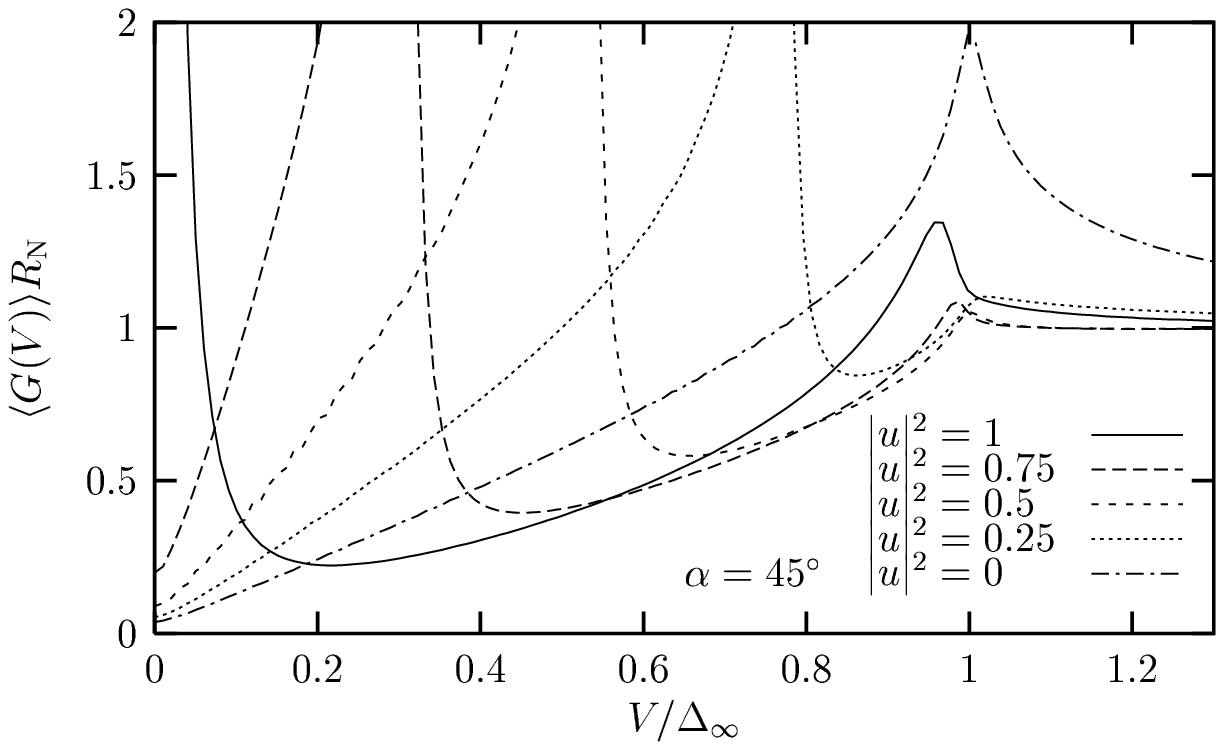}
\caption{Differential conductance for $m=2$ and $\alpha=0^\circ,45^\circ$; 
the specular scattering weight is varied.}\label{dom2}
\end{figure}

We study the simplest case $m=2$ with three mirrors with orientation
$\theta=0^\circ,\pm 45^\circ$, where two in- and two out-trajectories are
coupled. The $S$-matrix has the form
\begin{equation}\label{Smat-3fac}
{\bf S}=
\begin{pmatrix}
{\bf u}&  {\bf v}\\
{\bf v}&  {\bf u} 
\end{pmatrix}
\end{equation}
with ${\bf u}=u{\bf 1}_{n/2}$ and ${\bf v}=v{\bf 1}_{n/2}$;
the functions $u$ and $v$ are given by
\begin{equation}\label{S-3fac}
u(\tau)=\cos(\tau),\qquad v(\tau)=i\sin(\tau).
\end{equation}
For $\alpha=0^\circ$ and $|u|^2<1$ finite energy bound states occur; their
energies move to zero with decreasing weight of specular reflection. If
$|u|^2=0$ ($\tau=\pi/2$), i.e. no specular reflection is present, the bound
states reach zero energy. For $\alpha=45^\circ$ this is reversed:
non-specular reflection leads to a splitting of the zero energy bound states,
which grows with decreasing $|u|^2$ (Fig.~\ref{dom2}), as was discussed
qualitatively in~\cite{ShOz98}.
This model is in some aspects similar to a Josephson-contact with
a specular interface, where also a splitting of the zero energy bound state
can be observed.

\begin{figure}
\epsfig{file=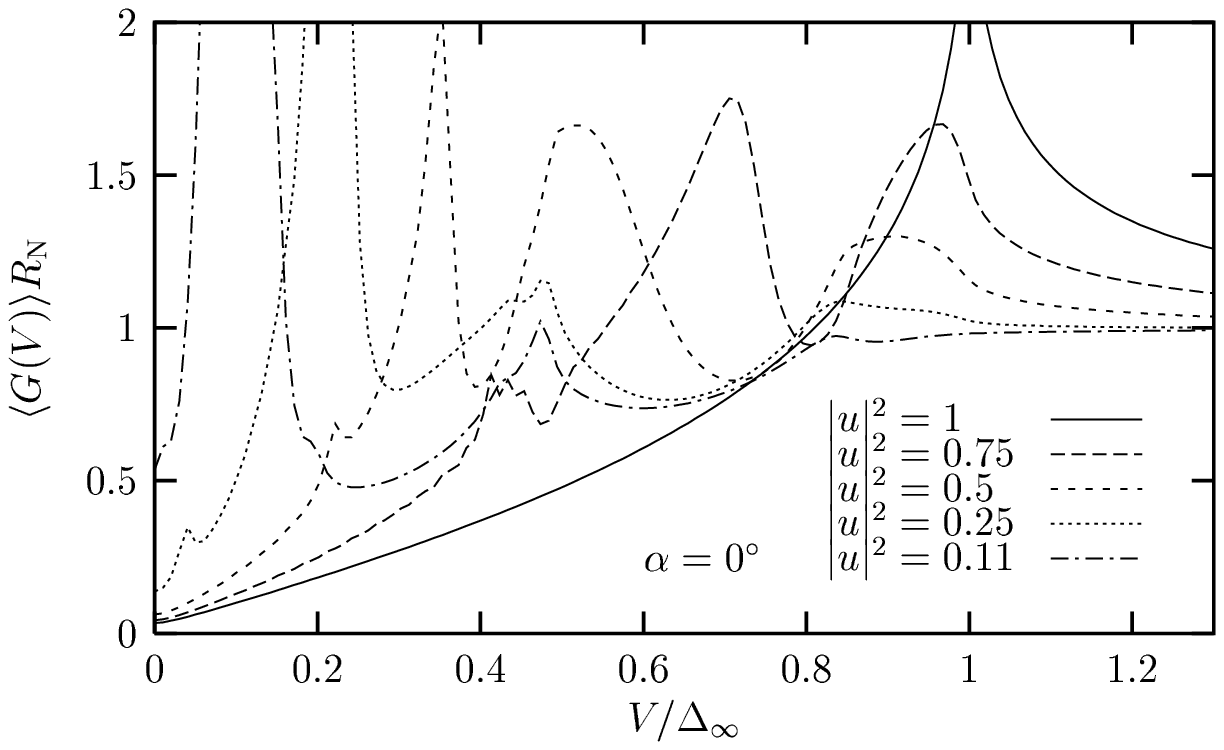}
\epsfig{file=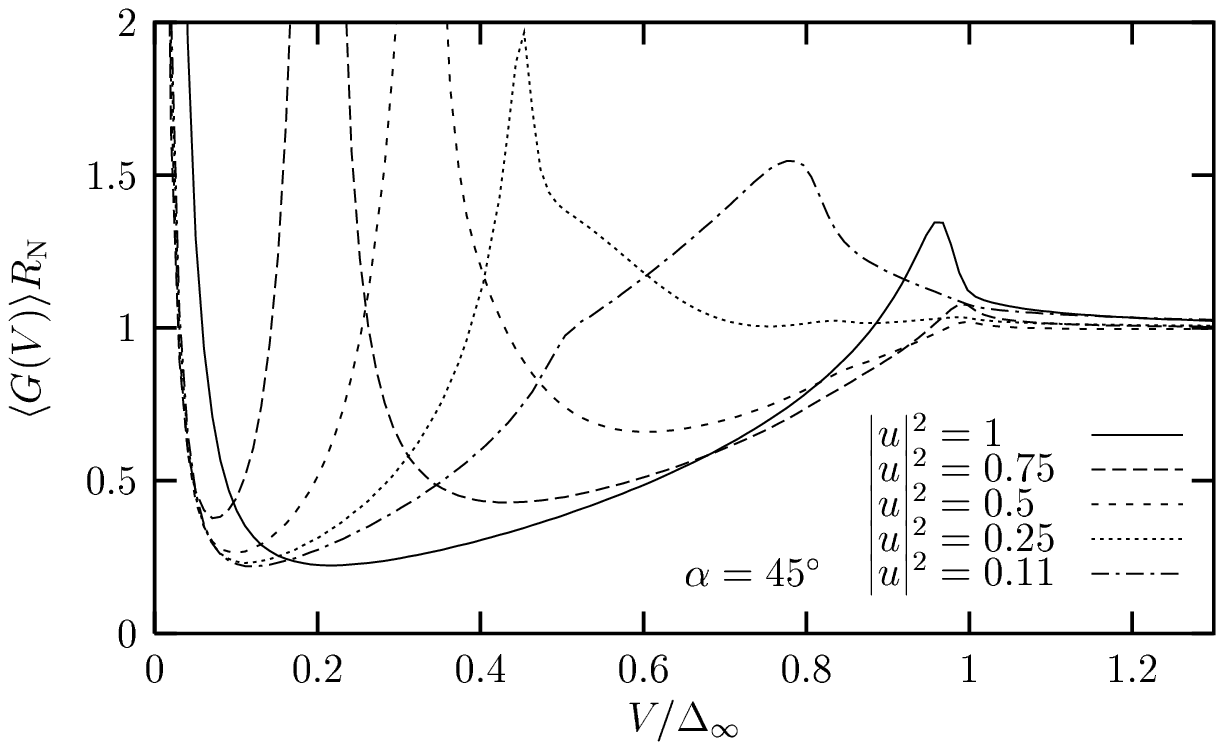}
\caption{Differential conductance for $m=3$ and $\alpha=0^\circ,45^\circ$; 
the specular scattering weight is varied.}\label{dom3}
\end{figure}

\begin{figure}
\epsfig{file=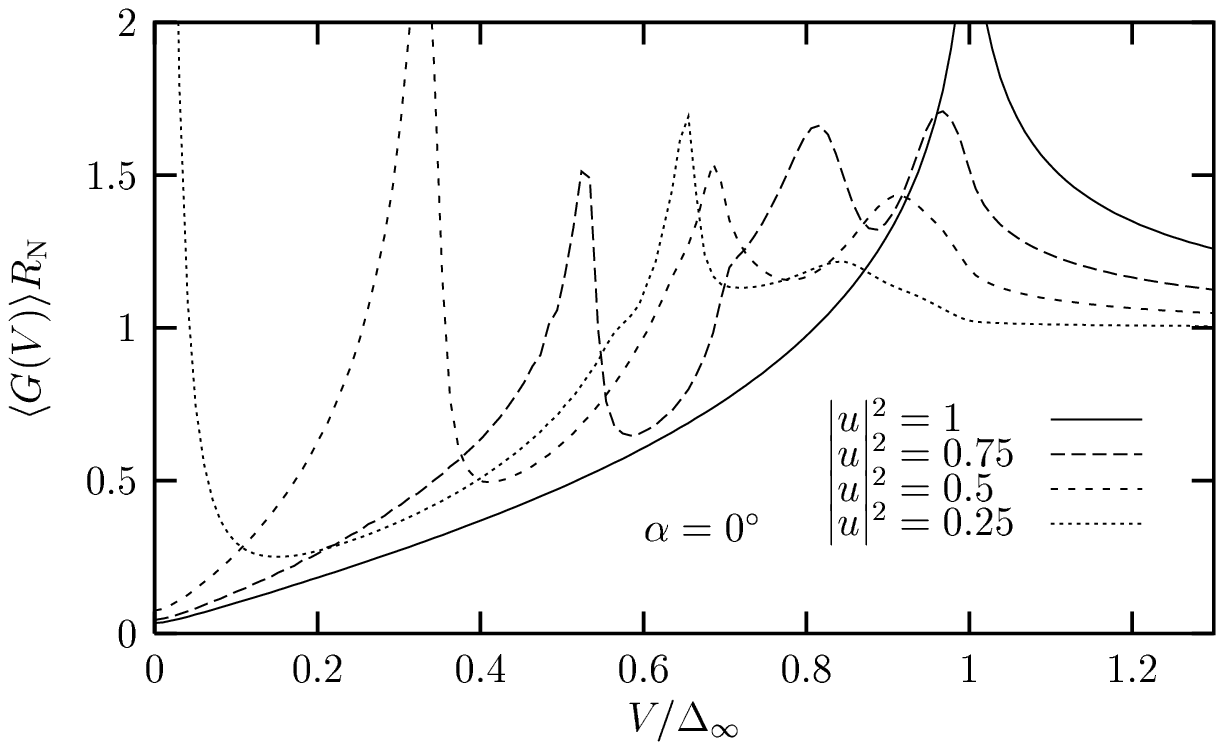}
\epsfig{file=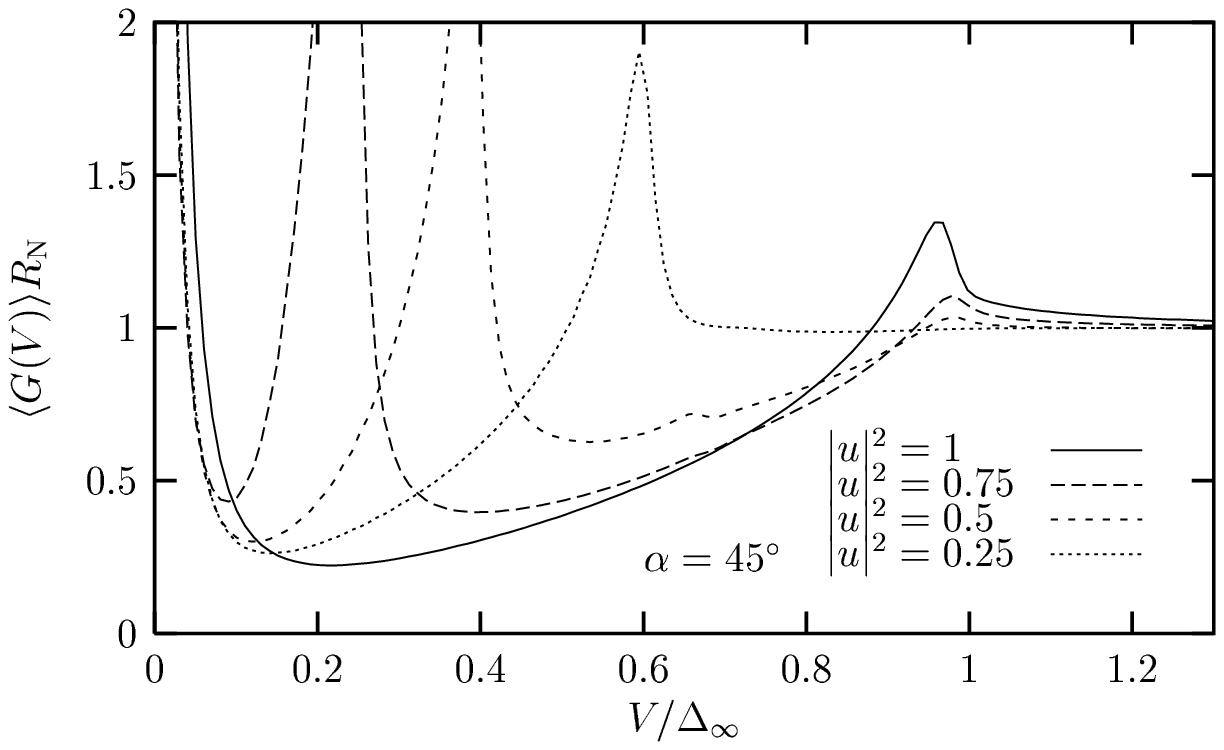}
\caption{Differential conductance for $m=4$ and $\alpha=0^\circ,45^\circ$; 
the specular scattering weight is varied.}\label{dom4}
\end{figure}

The case $m>2$, however, is more complex and cannot be mapped to any studied
model.
We consider the case $m=3$. The five mirrors have the relative orientations
$\theta=0^\circ,\pm 30^\circ,\pm 60^\circ$ and three in- and three
out-trajectories are connected; the $S$-matrix is given by
\begin{equation}\label{Smat-5fac}
{\bf S}=
\begin{pmatrix}
{\bf u} &  {\bf v}  & {\bf v}\\
{\bf v} &  {\bf u}  & {\bf v}\\
{\bf v} &  {\bf v}  & {\bf u}
\end{pmatrix}
\end{equation}
with ${\bf u}=u{\bf 1}_{n/3}$ and ${\bf v}=v{\bf 1}_{n/3}$.
As can be seen in Eq.~(\ref{scat-prob}) for $m=3$ it is only
possible to choose the specular scattering probability in the interval
$|u|^2\in[1/9,1]$.
For $\alpha=0^\circ$ non-specular scattering leads to finite energy bound
states which are moving to lower energies with decreasing $|u|^2$; but here
zero energy is not reached. In the case of $\alpha=45^\circ$
for $|u|^2<1$ one part of the zero energy bound state splits to finite
energies, whereas another part stays at zero energy with reduced spectral
weight (Fig.~\ref{dom3}).
In figs.~\ref{dom4} we also present the case $m=4$ with $7$ different mirrors
for the surface orientations $\alpha=0^\circ,45^\circ$.

Summing up our observations for the $\alpha=45^\circ$ case small
mirrors reduce the spectral weight of the zero energy bound states as a part
splits to finite energies. For the surface orientation $\alpha=0^\circ$ in
general (except some particular situations)  no zero energy bound state is
produced.

Finally we point out that a larger class of $S$-matrices can be used
to describe such surfaces
\begin{equation}\label{H-fac.gen}
H_{ij}=\sum\limits_{l=1}^{m-1}\tau_l\delta_{|i-j|,ln/m}.
\end{equation}
This model has almost the same properties as~(\ref{H-fac}), however, the
scattering probabilities for each mirror differ. Moreover it is possible to
combine the these mirrors with disorder just by multiplying the related
$S$-matrices and averaging as in section~\ref{RanSurf}.

\section{Discussion and Conclusion}

In the present paper, the scattering matrix approach has been
applied to describe $d$-wave superconductors in the vicinity of rough
surfaces. Two physical situations are examined: 
(i) a  surface with partially diffusive reflection described by random
scattering matrices; 
(ii) a surface with small tilted mirrors (facets) where the reflected wave is
a coherent mixture of waves propagating in several directions.

First, for the diffusive surface it appears that our results are very
similar to those found in other approaches such as the randomly
rippled wall model~\cite{Ya96} or  the thin dirty layer~\cite{Bu97}. 
In contrast to those calculations we treated the disorder by direct sampling.
We find that the deviation of the DOS from the average is rather small. 
Therefore, our calculations confirm the validity of
averaging procedures used in the afore-mentioned papers.  In
particular the broadening of the ZBCP due to increasing disorder is no
artifact of the approximate averaging. (Other models exist that show no
such broadening~\cite{Po99,Wa99}.)

Second, we studied a surface with tilted mirrors.  In contrast to the
model for large facets examined in~\cite{Fo97}, our model
describes a surface where faceting occurs on a scale small compared to
the coherence length.
These two models provide qualitatively different results: our model in
general shows no ZBCP for $\alpha=0$, whereas large facets lead to a
ZBCP for each surface orientation.

In experiments on high-$T_{\text{c}}$ materials, it
cannot be excluded that roughness on a scale of the coherence length or
larger occurs, which is beyond the model used in the current
paper. This might be the case in the experiment described in~\cite{Ap98},
where the width of the ZBCP is constant with varying disorder.
In some experiments~\cite{Co97,We98} a ZBCP
for $\alpha=0$ is observed, too, consistent with the model for large
facets~\cite{Fo97}.

\section*{Acknowledgments}

We would like to thank Y. Barash, M. Dzierzawa, M. Fogelstr\"om, and
J. Mannhart for helpful discussions. This work was supported in part by the
DAAD, the BMBF (project number 13N6918/1), and the Swedish Institute.

\end{document}